# Sorting OAM modes with metasurfaces based on raytracing improved optical coordinate transformation

ZHIBING LIU,[1] JIAHUI ZOU,[1] ZHAOYU LAI,[1] JIAJING TU,[2] SHECHENG GAO,[2,*] WEIPING LIU,[2] ZHAOHUI LI,[1,3]

[1]*State Key Laboratory of Optoelectronic Materials and Technologies, School of Electronics and Information Technology, Sun Yat-sen University, Guangzhou 510275, China*
[2]*Department of Electronic Engineering, College of Information Science and Technology, Jinan University, Guangzhou 510632, China*
[3]*Southern Marine Science and Engineering Guangdong Laboratory (Zhuhai), Zhuhai 519000, China*
*\*corresponding author: gaosc825@163.com*

**Abstract:** Optical coordinate transformation (OCT) has attracted widespread attention in the field of orbital angular momentum (OAM) (de)multiplexing or manipulation, but the performance of OCT would suffer from its distortion. In this paper, we quantitatively analyze the distortion of OCT from the perspective of ray optics, and explain its rationality to work under non-normal incident light. For the special case of log-polar coordinate transformation (LPCT), we use a raytracing assisted optimization scheme to improve its distortion, which is related to a Zernike polynomial based phase compensation. After raytracing optimization, the root mean square error (RMSE) of the focused rays is reduced to 1/5 of the original value and the physical optic simulation also shows great improvement. In the experiment, we use three phase masks which are realized by metasurfaces, the measured results show well consistency with the simulation. Results in this paper have great potential to improve the performance of OCT related applications.

## 1. Introduction

In recent years, the ever-growing demands for information bandwidth has posed a huge challenge to the current optical communication system. Since the time, frequency, wavelength and polarization dimensions of light have been deeply researched before, exploiting the space dimension of light, also known as space division multiplexing (SDM), is considered to be the next-generation communication technology [1,2]. Among SDM, one of the most representative subsets is orbital angular momentum (OAM) multiplexing, which have given rise to especial research interest in mode division multiplexing (MDM). Generally, light beam with spiral phase wavefront factor $e^{il\theta}$, where $\theta$ is the azimuthal coordinate, $l$ is the topological charge (TC), carry $l\hbar$ OAM per photon [3]. In principle, $l$ is an unbounded integer, OAM beams with different $l$ are orthogonal and thus have great application prospects in optical communication [4]. Apart from communication, OAM beams have also been applied to fields such as optical manipulation [5], optical tweezer [6], imaging [7], and quantum information [8]. In all of these applications, $l$ is a determined parameter since it is directly related to the quantity of OAM. Owing to its great importance, measuring $l$ or demultiplexing different OAM beams have attracted extensive researches [9-15].

Although there have been many studies on OAM measurement, simultaneously demultiplexing multiple OAM beams without shrinkage of efficiency is still challenging. To cope with this problem, in 2010, Berkhout et al. proposed a scheme which is based on optical coordinate transformation (OCT). In their way, annular OAM beam with spiral phase is firstly transformed into rectangular beam with plane wave phase and then can be sorted by a focused lens, this method also known as log-polar coordinate transformation (LPCT) [16]. Due to the simple implementation and unified efficiency, the LPCT method has received widespread concerns [17-22]. Inspired by the LPCT, some other versions of OCT for OAM manipulation have also been proposed, such as circular-sector transform [23] and spiral transform [24]. Though the OCT has its great advantage in sorting OAM, its performance could suffer from severe distortion when used for higher-order OAM state, this is because the beam need to assumed to be normally incident while the higher-order OAM beam has large skew angle [16]. To relieve this problem, Lavery et al. have analyzed the distortion and used two large phase masks to achieve up to 28 orders of OAM measurement in 632nm [19]. However, though we know the distortion could be introduced by non-normal incident light rays, there are still no quantitively analysis of the distortion and no corresponding solutions to improve the performance of the OAM sorter that work under specific configuration.

In this work, through analyzing the ray's trajectory among the coordinate transformation process, we will explain the rationality to use the OCT beyond normal incident assumption. It should be noted that the ray deviated from normal incidence will go through incorrect deflection in the second phase plate. Then we can further provide a quantitative result of the distortion in OCT, and for the case of LPCT, we give an upper bound of $l$ that can be measured under certain distortion. In addition, to alleviate the performance degradation caused by the distortion, we present a scheme to improve the distortion by raytracing assisted optimization, which is related to Zernike polynomial based phase compensation. After the optimization, we use three phase masks realized by metasurfaces to build the experimental setup and implement the measurement.

## 2. Theoretical analysis of distortion

Generally, the OCT maps the points of two separate planes, let $(x, y)$ denote the first plane and $(u, v)$ denote the second. Assuming light is normally incident, to achieve the map, from the view of ray optics and the paraxial approximation condition, a phase mask $\varphi$ should be put on the first plane and its gradient satisfied

$$\frac{\partial \varphi}{\partial x} = \frac{2\pi}{\lambda} \frac{u-x}{f}, \quad \frac{\partial \varphi}{\partial y} = \frac{2\pi}{\lambda} \frac{v-y}{f}, \quad (1)$$

where $\lambda$ is the wavelength, and $f$ is the distance between the two planes. When the ray arrives at the second plane, similarly, we need a phase mask $\psi$ to compensate the ray's deflection, and its gradient satisfied

$$\frac{\partial \psi}{\partial u} = \frac{2\pi}{\lambda} \frac{x-u}{f}, \quad \frac{\partial \psi}{\partial v} = \frac{2\pi}{\lambda} \frac{y-v}{f}. \quad (2)$$

In general, given a reasonable relationship between $(u, v)$ and $(x, y)$, we can figure out the corresponding $\varphi$ and $\psi$ from above equations to achieve the transformation. As shown in Fig. 1, then a ray normally incident on the $(x, y)$ in the first mask, will pass through the device and arrives at the $(u, v)$, and normally incident out from the second mask. However, if the ray is non-normally incident, its trace will deviate from the position $(u, v)$, and also experience unexpected phase gradients. In this situation, further discussion is needed.

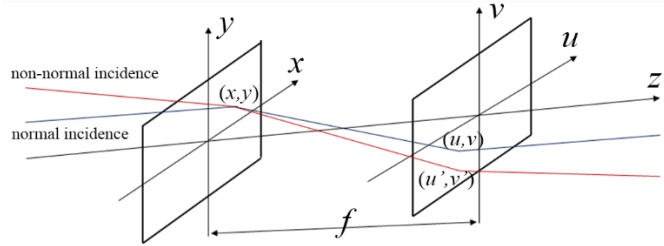

Fig. 1. Coordinate transformation with normal incidence and non-normal incidence.

For simplicity, assuming the incident ray can be described by phase distribution $P$, then the ray's deviation in the second mask caused by its phase gradient can be described as

$$\vec{d} = (d_u, d_v) = \frac{\lambda f}{2\pi} \nabla P. \tag{3}$$

This means that if the incident light is non-normal incidence or $\nabla P \neq 0$, it will cause the distortion of position of OCT. Since OAM beam has skew angle, to deal with this problem for better performance, $|\vec{d}| \ll r$ is supposed in reported works [17, 24] and thus can cause a restriction on the $l$ of OAM beam with given size $r$.

In above, the position after transformation is considered within the scope of ray optics. However, for the application of OAM (de)multiplexing or manipulation, the distortion of position after OCT is not the only problem, because the phase transformation is the main purpose. To better evaluate the distortion, ray's direction after OCT is need to be considered. Due to the position distortion, the ray will obtain a corresponding phase gradient change in the second plane, and by using Eq. (2) the gradient change can be described as

$$\Delta \frac{\partial \psi}{\partial u} = \frac{2\pi}{\lambda f}(-d_u + \Delta x), \quad \Delta \frac{\partial \psi}{\partial v} = \frac{2\pi}{\lambda f}(-d_v + \Delta y), \tag{4}$$

where $\Delta x = x(u+d_u, v+d_v) - x(u,v)$, $\Delta y = y(u+d_u, v+d_v) - y(u,v)$, since the ray's deflection caused by the first mask can be offset by the second mask, its ultimately phase gradient accumulated by the transformation process can be expressed as

$$\vec{G} = (G_x, G_y) = \nabla P + \left(\Delta \frac{\partial \psi}{\partial u}, \Delta \frac{\partial \psi}{\partial v}\right) = \frac{2\pi}{\lambda f}(\Delta x, \Delta y). \tag{5}$$

Eq. (5) shows the phase gradient that represents the ray's direction, to explore the ray's property, we perform Taylor expansion on $\vec{G}$ and the gradient can be written as

$$\vec{G} = \frac{2\pi}{\lambda f}\left(\vec{d}\cdot\nabla'x, \vec{d}\cdot\nabla'y\right) + \Delta\vec{G}, \tag{6}$$

where $\nabla'$ is arithmetic of $(u,v)$ coordinate, $\Delta\vec{G}$ is higher order residuals. From Eq. (2) and using $\partial x/\partial v = \partial y/\partial u$, the ray's phase gradient can be further written as

$$\vec{G} = \nabla' P + \Delta\vec{G}. \tag{7}$$

From the right hand of Eq. (7), we can see that the input ray with phase gradient $\nabla P$ has transformed into the output ray with gradient $\nabla' P$ after OCT, it is also mean that the phase distribution in the first plane has transformed into the second plane. In this way, the rationality of OCT to transform the phase distribution is justified by the first-order Taylor approximation. Meanwhile, the residual term in the equation is actually a distortion that cause the diffusion of the ray and can be calculated from Eq. (5) and Eq. (7). For convenience, this distortion can also be approximated by the second order Taylor expansion

$$\Delta\vec{G} \approx \frac{\pi}{\lambda f}\vec{d}\vec{d} : \left[(\nabla'\nabla'x)\vec{i} + (\nabla'\nabla'y)\vec{j}\right]. \tag{8}$$

In most situations, the coordinate transform is conformal, or more specifically, it is a conjugate analytic function

$$x + iy = F(u - iv), \tag{9}$$

where $F$ is an analytic function, then using $F$ the distortion can also be expressed as

$$\Delta G_x + i\Delta G_y = \frac{\pi}{\lambda f}(d_u - id_v)^2 F^{(2)}(u - iv). \tag{10}$$

From Eq. (10), we can conveniently calculate the distortion of various OCT. From Eq. (3) we can see the position deviation is proportional to $|\vec{d}|$, while here the phase gradient distortion $|\Delta\vec{G}|$ is proportional to $|\vec{d}|^2$.

Next, we start to consider the special case of LPCT that have widely used for OAM sorting. In this case, $u = -a\ln(r/b)$, $v = a\theta$, where $a$, $b$ are positive parameters of the transformation, then we can derive the corresponding relationship

$$x + iy = be^{-(u-iv)/a}. \tag{11}$$

In addition, for an OAM beam with given radius $r$ and phase wavefront $P = l\theta$, from Eq. (3) the deviation caused by its skew angle can be described as

$$\vec{d} = \frac{\lambda f}{2\pi}\nabla(l\theta) = \frac{\lambda fl}{2\pi r}(-\sin\theta, \cos\theta). \tag{12}$$

Then from Eq. (10), the distortion can be expressed as

$$\Delta\vec{G} = \frac{\lambda fl^2}{4\pi ra^2}(-\cos\theta, \sin\theta). \tag{13}$$

Eq. (13) is the distortion of LPCT, if we postulate the distortion is smaller than the difference between adjacent OAM modes, because otherwise the crosstalk will become too large, that's mean $|\Delta\vec{G}|/(1/a) \leq 1$, and we can get

$$|l| \leq \sqrt{4\pi ar/\lambda f}. \tag{14}$$

This is the upper bound of the OAM states can be sorted by the LPCT under given distortion. For other type coordinate transformations such as spiral transformation [24], through same treatment, similar constraint can also be obtained.

### 3. Raytracing assisted distortion improvement

In order to improve the distortion of LPCT derived in above section, we introduce the raytracing to simulate the process. Since the raytracing is a common tool for geometrical optics design, it should also be suitable for analyzing OCT. Here, to implement the LPCT, we use three phase masks to work as transformer, compensator, and a lens respectively to separate or sort the input OAM beams. These three phase masks are shown in Fig. 2(a-c). In addition, the input OAM beam used here is LG beam, of which beam size $r$ is varied according to $l$ and follow $r = w\sqrt{|l|}$, where $w$ is the waist radius of the gauss mode. In this situation, to limit the interference on adjacent order as derived in above section, the range of TC should satisfy

$$|l| \leq \sqrt[3]{(4\pi aw/\lambda f)^2}. \tag{15}$$

The raytracing then is done according to Eq. (1-2) where paraxial condition is used. Fig. 2(d) shows the raytracing schematic diagram of sorting multiple OAM beams, different OAM beam has corresponding skew angle determined by its order and radius, and the rays will go through the three masks and convergence at the focal plane. As is shown in Fig. 2(e), the rays of OAM beams with different order $l$, where $-10 \leq l \leq 10$, have different radius. The waist $w$ of the LG beam is set to be $0.18\ mm$, and the parameters of the LPCT respectively set as, $\lambda =$ 1.55 $\mu m$, $a = 1.5/2\pi\ mm$, $b = 0.45\ mm$, and $f = 15\ mm$. For simplicity, the focal length of the convergence lens is also set to be $f$. Then from Eq. (15) we can find the critical TC is about $l = 8$. Accordingly, Fig. 2(f) shows the results of focused rays based on the installation, it can be seen that the ray will go to adjacent focus when the order is larger than 8. This ray optics simulation result confirmed our theoretical analysis and shows good accuracy to approximate the distortion of LPCT using second order Taylor expansion.

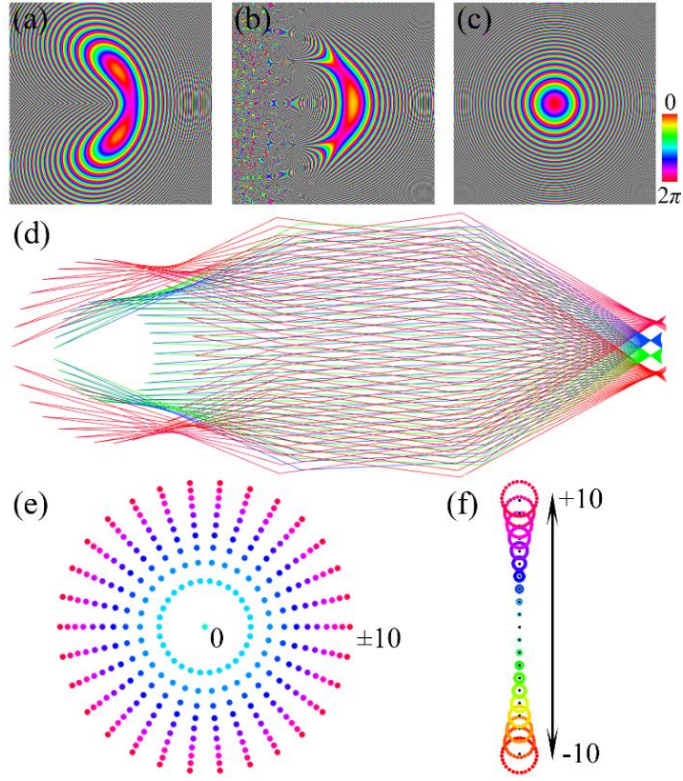

Fig. 2. Raytracing simulation of LPCT. (a)-(c) The three phase masks used in the raytracing simulation. (d) Schematic diagram of raytracing to simulating LPCT. (e) The start points of the rays in the input plane and (f) The endpoints of rays in the focal plane.

Next, to overcome the huge distortion of LPCT resulted by the skew angle of input OAM beam, we further use the raytracing to optimize the phase masks. To implement the optimization, we define a Zernike polynomial based phase compensation. For every mask, the first 45 order Zernike polynomials are added to express the phase compensation, fewer terms of polynomials may lead to worse results while more terms would increase the difficulty of optimization. The modified phase mask thus can be described as follow

$$Q'(x,y) = Q(x,y) + \sum_{k=1}^{45} a_k Z_k(x,y), \qquad (16)$$

where $Q$ is the original phase, $Z_k$ is the $k$-order Zernike polynomial, $a_k$ is the corresponding coefficient, and $Q'$ is the modified phase. When a ray goes through the mask, its traces will be regulated by extra phase denoted by Zernike polynomial, and the coefficients $a_k$ thus can be used as optimizing variables. Besides, we define the root mean square error (RMSE) as the extent of distortion for the optimization, this can be denoted by

$$FoM = \sqrt{\frac{1}{N}\sum_{n=1}^{N}\left[(x_n - \tilde{x}_n)^2 + (y_n - \tilde{y}_n)^2\right]}, \qquad (17)$$

where $FoM$ needs to be minimized, $N$ is the number of rays, $(x, y)$ and $(\tilde{x}, \tilde{y})$ are the ray's arrival position and ideal destination in the focal plane. After set up this raytracing model, we use an evolutionary algorithm named covariance matrix adaptive moment estimation (CMA-ES) to regulate the optimizing parameters or the Zernike coefficients [25], which is a commonly used global optimization algorithm. Other methods such as genetic algorithm (GA) [26], or particle swarm optimization (PSO) [27] may be also valid here.

Figs. 3(a-c) show the phase masks after Zernike polynomial based phase compensation. Fig. 3(d) shows the corresponding convergence of rays in focal plane. There is a great deal of diminution of distortion especially for the higher-order OAM, and the RMSE have dropped from original value of $15.87\mu m$ to $3.27\mu m$ while the spacing of adjacent TC is $15.50\mu m$. In Figs. 3(e-f) and Figs. 3(g-h), we show the physical optics based simulation results before and after optimization. Fig. 3(e) and 3(f) show the intensity distribution of multiple OAM beams in the focal plane and corresponding normalized intensity distribution along the central axis. From these focused spots we can find that with the increasing $l$, there are also increasing distortion. The critical order is $l = 8$ when the distortion goes to heavily influence the adjacent spots, which is in good agreement with the theoretical prediction. Correspondingly, the simulation results after optimization are shown in Fig. 3(g) and 3(h). From the figures we can see a huge improvement of the sorting performance. The spots in Fig .3(g) have roughly uniform size and less overlap with adjacent spots. Meanwhile, the intensity distribution of the sorted OAM beams along the central axis in Fig. 3(h) also indicate less interference to other beams. These simulation results signify that the distortion improvement is effective.

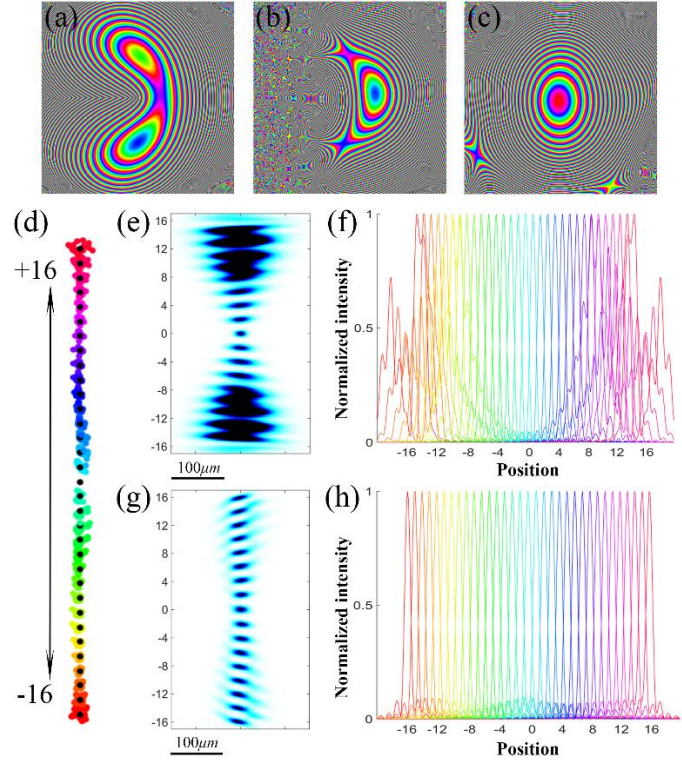

Fig. 3. Simulation results after raytracing optimization. (a)-(c) The phase masks after raytracing assisted optimization. (d) The ray's arrival position in the focal plane after optimization. (e-f) Sorting results before optimization. (g-h) Sorting results after optimization.

## 4. Device preparation and characterization

In order to prepare above optimized phase masks for experiment, we use three metasurfaces to provide the corresponding phase modulation. Here the function of metasurfaces is mainly concentrate on $2\pi$ phase coverage and high transmittance, a most favorable selection is the dielectric metasurface that have demonstrated in reported researches [28-30]. As shown in Fig. 4(a), the metaunit of our dielectric metasurface is composed of amorphous silicon (a-Si) cylinder and $SiO_2$ substrate, the metaunit thus is polarization insensitive. In addition, to prevent the cylinder from collapsing in the fabrication, 20nm a-Si is remained in the substrate. Under the given structure, to maximize the transmittance while cover $2\pi$ phase, we have done a parameter sweep to determine the metaunit and respectively set $p = 760nm$ and $h = 890nm$ according to the sweeping results. The phase and transmittance response of the metaunit are shown in Fig. 4(b). We select 96 structures with phase difference $2\pi/96$ through interpolation and have marked them with asterisk in the Fig. 4(b). Based on the metaunit, coverage of $2\pi$ phase can be achieved and the average transmittance of these structures is 90.77%.

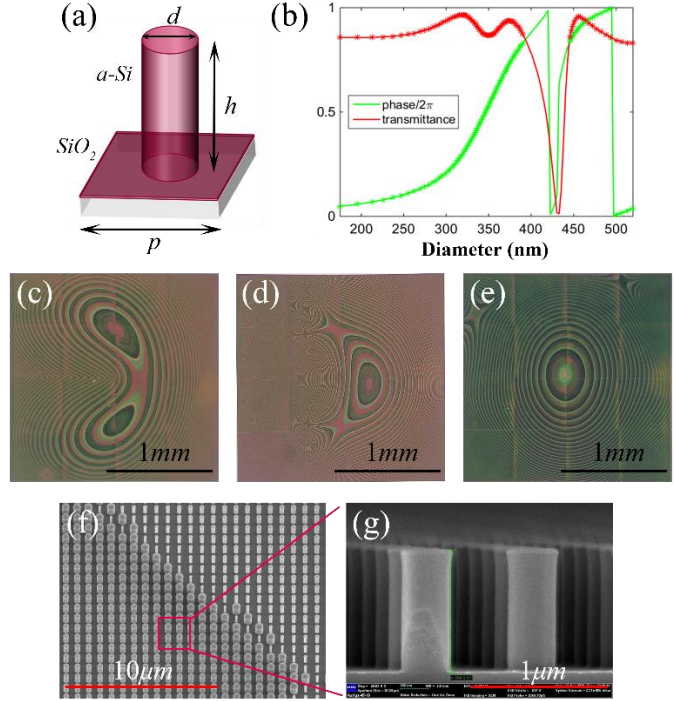

Fig. 4. Metasurfaces design and fabrication. (a) The schematic diagram of metaunit. (b) The phase and transmittance response of the metaunit. (c-e) The fabricated metasurfaces that are captured in microscope. (f-g) The partial fabrication details of the device in SEM.

After the metaunit design, we generate the layout to meet the required phase masks in above section and fabricate the device accordingly. The fabricated process can be briefly described as follow. Firstly, a layer of 890-nm-thick a-Si is evaporated in the SiO$_2$ substrate using chemical vapor deposition (CVD). After pretreatment of the sample, the AR-P6200.13 (ARP) resist is spin coated in the a-Si film. Before electron beam lithography (EBL), a layer of Al is deposit in the ARP resist. The device then is put into the EBL device to write the mask. After exposure, we remove the Al layer and develop the resist to get the patterned mask. Afterwards, we use inductive coupled plasma (ICP) to etch the a-Si layer. Finally, the ARP resist is removed and the device is fabricated. In Figs. 4(c-e), we show the fabricated metasurfaces that are captured in microscope, which are consistent with Figs. 2(a-c). Figs. 4(f-g) show the partial fabrication detail in scanning electron microscope (SEM). From these images we can see the device is well fabricated and the pillars have been etched appropriately.

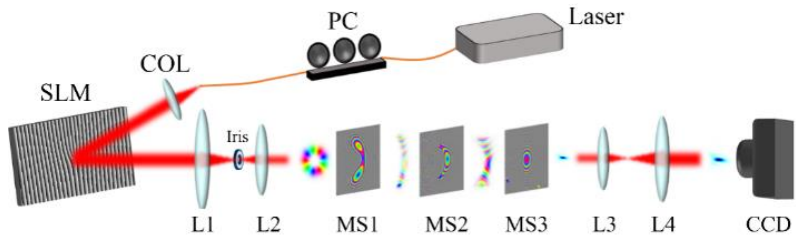

Fig. 5. Experiment setup. PC: polarization controller. COL: collimator. SLM: space light modulator. L1-L4: the lens with focal length, f1 = 15cm, f2 = 7.5cm, f3 = 5cm, f4 = 12.5cm. MS1-MS3: the fabricated metasurfaces. CCD: charge coupled device.

Based on the metasurfaces described above, we built the experiment measurement setup as shown in Fig. 5. A laser with 1550nm wavelength is used as the light source and the generated beam is collimated into SLM, where we use a PC to regulate its polarization. After the reflection and modulation of the SLM, the OAM-carrying beam then is contracted by a pair of lenses L1 and L2 to match the size of the metasurfaces. To diminish the unexpected diffraction order of SLM, an iris is placed between the two lenses. The OAM beam then arrive at the metasurface MS1 and undergo the modulation of MS1-3. Finally, since the focal spot is small, we use lenses L3 and L4 to expand it and the CCD is used to detect the results.

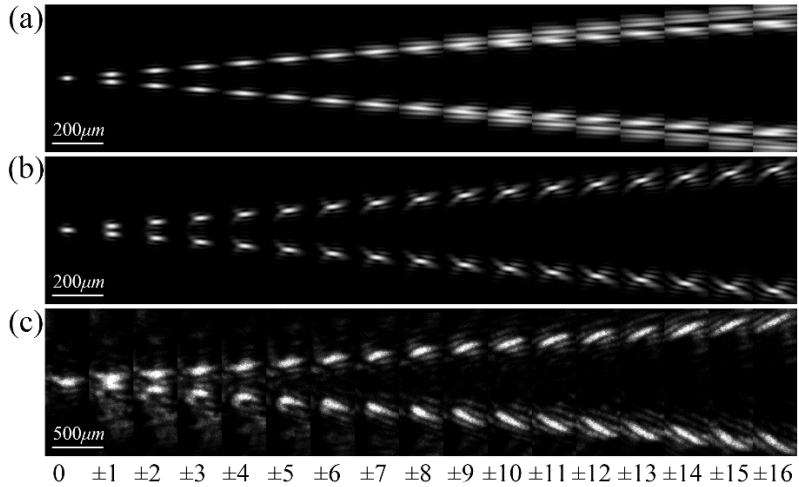

Fig. 6. Sorting results of OAM beams with -16≤ $l$ ≤16. (a) Simulation results before raytracing optimization. (b) Simulation results after raytracing assisted optimization. (c) The experimental detection results.

At last, we carry out the experiment, the sorting results of OAM beams with $-16 \leq l \leq 16$ are shown in Fig. 6. As comparison, Figs. 6(a-b) show the simulation results before and after optimization. It can be seen that the simulation results have severe distortion before optimization when $l \geq 8$, but after optimization, the divergence of focal spot are improved significantly. As shown in Fig. 6(c), the experimental detection of OAM beams up to $\pm 16$ order is achieved, and the distortion of higher-order OAM beams also have been significantly suppressed. Further, in order to quantify the effect of distortion improvement, the detected transfer matrix of simulation and experiment are shown in Fig. 7. Since the focal spot is slightly inclined, to better evaluate the improvement of the spot divergence, the relative power of the detected transfer matrix is determined by mean intensity along the vertical axis in focal plane. In this way, from Fig. 7(a) we can see that the crosstalk is very high before optimization, and the higher -order OAM beams become gradually indistinguishable with adjacent beams when $l \geq 8$. This situation can be improved by raytracing assisted optimization, as shown in Figs. 7(b-c), both the simulation and experimental results show well distinction even for $l= \pm 16$. Though there are some differences between the simulation and experiment results, which may be due to the manufacturing errors and the imperfect experimental setup. On the whole, the performance of LPCT has been improved greatly, and the experimental results can well support the reliability of simulation and indicate a good effectiveness of the raytracing assisted distortion improvement.

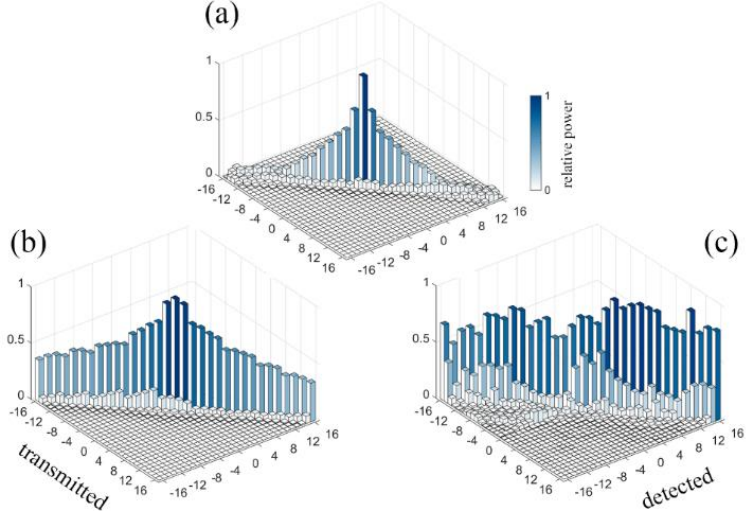

Fig. 7. The detected transfer matrices of the OAM modes sorter. (a) Simulation before raytracing assisted optimization. (b) Simulation results after optimization. (c) Experimental results.

## 5. Conclusions

In conclusion, we have provided a quantitative analysis of the distortion in OCT, and also clarified the rationality of optical coordinate transformation for non-normally incident beams in the view of geometrical optics. For the case of LPCT, we give an upper bound of $l$ that can be measured under certain distortion. As the OCT can only work well in the first order Taylor expansion approximation, the sorting distortion will increases as the TC of OAM beam increases. Therefore, to alleviate this problem, we further provide a method based on the raytracing and Zernike polynomial based phase compensation. After raytracing optimization, the RMSE of the focused rays can be reduced to 1/5 of the original and the physical optic simulation also shows obviously distortion improvement. At last, to confirm the simulation results, we implement the experiment using three fabricated metasurfaces. The experimental results are in good agreement with the simulation and show obvious performance improvement after raytracing optimization. Results in this paper may be also help to further improve other OCT related applications and have potential to extend by adding more masks.

**Funding**

National Key Research and Development Program of China (2019YFA0706300); National Natural Science Foundation of China (U2001601, 62035018, 61775085, 61875076, 61935013, U1701661); Guangzhou Science and Technology Program key projects (201904020048); Local Innovation and Research Teams Project of Guangdong Pearl River Talents Program (2017BT01X121); Science and Technology Planning Project of Guangdong Province (2017B010123005, 2018B010114002); Leading Talents Program of Guangdong Province (00201502); State Key Laboratory on Integrated Optoelectronics (IOSKL2019KF12).

**Disclosures**

The authors declare no conflicts of interest.

## References

1. G. Li, N. Bai, N. Zhao, and C. Xia, "Space-division multiplexing: the next frontier in optical communication," Adv. Opt. Photonics 6, 413 (2014).
2. D. J. Richardson, J. M. Fini, and L. E. Nelson, "Space-division multiplexing in optical fibres," Nat. Photonics 7, 354–362 (2013).
3. L. Allen, M. W. Beijersbergen, R. J. C. Spreeuw, and J. P. Woerdman, "Orbital angular momentum of light and the transformation of Laguerre-Gaussian laser modes," Opt. Angular Momentum 45, 31–35 (2016).
4. A. E. Willner, H. Huang, Y. Yan, Y. Ren, N. Ahmed, G. Xie, C. Bao, L. Li, Y. Cao, Z. Zhao, J. Wang, M. P. J. Lavery, M. Tur, S. Ramachandran, A. F. Molisch, N. Ashrafi, and S. Ashrafi, "Optical communications using orbital angular momentum beams," Adv. Opt. Photonics 7, 66 (2015).
5. K. Dholakia and T. Čižmár, "Shaping the future of manipulation," Nat. Photonics 5, 335–342 (2011).
6. M. Padgett and R. Bowman, "Tweezers with a twist," Nat. Photonics 5, 343–348 (2011).
7. A. Jesacher, M. Ritsch-Marte, and R. Piestun, "Three-dimensional information from two-dimensional scans: a scanning microscope with postacquisition refocusing capability," Optica 2, 210 (2015).
8. D. S. Ding, W. Zhang, Z. Y. Zhou, S. Shi, G. Y. Xiang, X. S. Wang, Y. K. Jiang, B. Sen Shi, and G. C. Guo, "Quantum storage of orbital angular momentum entanglement in an atomic ensemble," Phys. Rev. Lett. 114, 1–5 (2015).
9. J. Leach, M. J. Padgett, S. M. Barnett, S. Franke-Arnold, and J. Courtial, "Measuring the Orbital Angular Momentum of a Single Photon," Phys. Rev. Lett. 88, 4 (2002).
10. G. Gibson, J. Courtial, M. J. Padgett, M. Vasnetsov, V. Pas'ko, S. M. Barnett, and S. Franke-Arnold, "Free-space information transfer using light beams carrying orbital angular momentum," Opt. Express 12, 5448 (2004).
11. I. Moreno, J. A. Davis, D. M. Cottrell, N. Zhang, and X.-C. Yuan, "Encoding generalized phase functions on Dammann gratings," Opt. Lett. 35, 1536 (2010).
12. J. M. Hickmann, E. J. S. Fonseca, W. C. Soares, and S. Chávez-Cerda, "Unveiling a truncated optical lattice associated with a triangular aperture using light's orbital angular momentum," Phys. Rev. Lett. 105, 1–4 (2010).
13. H. L. Zhou, D. Z. Fu, J. J. Dong, P. Zhang, D. X. Chen, X. L. Cai, F. L. Li, and X. L. Zhang, "Orbital angular momentum complex spectrum analyzer for vortex light based on the rotational Doppler effect," Light Sci. Appl. 6, (2017).
14. N. K. Fontaine, R. Ryf, H. Chen, D. T. Neilson, K. Kim, and J. Carpenter, "Laguerre-Gaussian mode sorter," Nat. Commun. 10, 1–7 (2019).
15. Z. Liu, S. Yan, H. Liu, and X. Chen, "Superhigh-Resolution Recognition of Optical Vortex Modes Assisted by a Deep-Learning Method," Phys. Rev. Lett. 123, 183902 (2019).
16. G. C. G. Berkhout, M. P. J. Lavery, J. Courtial, M. W. Beijersbergen, and M. J. Padgett, "Efficient sorting of orbital angular momentum states of light," Phys. Rev. Lett. 105, 8–11 (2010).
17. M. P. J. Lavery, D. J. Robertson, G. C. G. Berkhout, G. D. Love, M. J. Padgett, and J. Courtial, "Refractive elements for the measurement of the orbital angular momentum of a single photon," Opt. Express 20, 2110 (2012).
18. M. Mirhosseini, M. Malik, Z. Shi, and R. W. Boyd, "Efficient separation of the orbital angular momentum eigenstates of light," Nat. Commun. 4, 1–6 (2013).
19. M. P. J. Lavery, D. J. Robertson, A. Sponselli, J. Courtial, N. K. Steinhoff, G. A. Tyler, A. E. Wilner, and M. J. Padgett, "Efficient measurement of an optical orbital-angular-momentum spectrum comprising more than 50 states," New J. Phys. 15, (2013).
20. G. Ruffato, M. Massari, and F. Romanato, "Diffractive optics for combined spatial-and mode-division demultiplexing of optical vortices: Design, fabrication and optical characterization," Sci. Rep. 6, 1–12 (2016).
21. C. Wan, J. Chen, and Q. Zhan, "Compact and high-resolution optical orbital angular momentum sorter," APL Photonics 2, (2017).
22. G. Ruffato, M. Massari, M. Girardi, G. Parisi, M. Zontini, and F. Romanato, "Non-paraxial design and fabrication of a compact OAM sorter in the telecom infrared," Opt. Express 27, 24123 (2019).
23. Y. Wen, I. Chremmos, Y. Chen, J. Zhu, Y. Zhang, and S. Yu, "Spiral Transformation for High-Resolution and Efficient Sorting of Optical Vortex Modes," Phys. Rev. Lett. 120, 1–2 (2018).
24. G. Ruffato, M. Massari, and F. Romanato, "Multiplication and division of the orbital angular momentum of light with diffractive transformation optics," Light Sci. Appl. 8, (2019).
25. N. Hansen, S. D. Müller, and P. Koumoutsakos, "Reducing the time complexity of the derandomized evolution strategy with covariance matrix adaptation (CMA-ES)," Evol. Comput. 11, 1–18 (2003).
26. S. Forrest, "Genetic algorithms," Comput. Sci. Handbook, Second Ed. 14-1-14–15 (2004).
27. M. Clerc, "Particle Swarm Optimization," Part. Swarm Optim. 1942–1948 (2010).
28. A. Arbabi, Y. Horie, M. Bagheri, and A. Faraon, "Dielectric metasurfaces for complete control of phase and polarization with subwavelength spatial resolution and high transmission," Nat. Nanotechnol. 10(11), 937–943 (2015).
29. A. Arbabi, Y. Horie, A. J. Ball, M. Bagheri, and A. Faraon, "Subwavelength-thick lenses with high numerical apertures and large efficiency based on high-contrast transmitarrays," Nat. Commun. 6, 2–7 (2015).
30. B. A. W. Ang, Y. U. W. En, J. I. Z. Hu, Y. Ujie, C. Hen, and S. Y. U. Iyuan, "Sorting full angular momentum states with Pancharatnam-Berry metasurfaces based on spiral transformation," 28, 16342–16351 (2020).